\documentclass[11pt,twoside]{article}
\usepackage{asp2014}

\aspSuppressVolSlug
\resetcounters

\begin{document}

\title{European Virtual Observatory Schools}

\author{Fran~Jim\'{e}nez-Esteban$^1$, Mark~Allen$^2$, Stefania~Amodeo$^2$, Miriam~Cort\'{e}s-Contreras$^1$, Sebastien~Derriere$^2$, Hendrik~Heinl$^2$, Ada~Nebot$^2$, and Enrique~Solano$^1$}
\affil{$^1$Centro de Astrobiolog\'{i}a (CSIC-INTA), E-28692 Villanueva de la Ca\~{n}ada, Madrid, Spain; \email{fran.jimenez-esteban@cab.inta-csic.es}}
\affil{$^2$Universit\'{e} de Strasbourg, CNRS, Observatoire astronomique de Strasbourg, UMR 7550, F-67000 Strasbourg, France}

\paperauthor{F.~Jim\'{e}nez-Esteban}{fran.jimenez-esteban@cab.inta-csic.es}{0000-0002-6985-9476}{Centro de Astrobiolog\'{i}a (CSIC-INTA)}{Astrophysics Department}{Villanueva de la Ca\~{n}ada}{Madrid}{E-28692}{Spain}
\paperauthor{Sample~Author2}{Author2Email@email.edu}{ORCID_Or_Blank}{Author2 Institution}{Author2 Department}{City}{State/Province}{Postal Code}{Country}



  
\begin{abstract}
The European Virtual Observatory (VO) initiative organises regular VO
schools since 2008. The goals are twofold: i) to expose early-career
European astronomers to the variety of currently available VO tools
and services so that they can use them efficiently for their own
research and; ii) to gather their feedback on the VO tools and
services and the school itself. During the schools, VO experts guide
participants on the usage of the tools through a series of predefined
real science cases, an activity that took most of the allocated
time. Participants also have the opportunity to develop their own
science cases under the guidance of VO tutors. These schools have
demonstrated to be very useful for students, since they declare to
regularly use the VO tools in their research afterwards, and for us,
since we have first hand information about the user needs. Here, we
introduce our VO schools, the approach we follow, and present
the training materials that we have developed along the years.
\end{abstract}

\section{Introduction}

The European Virtual Observatory (Euro-VO; https://www.euro-vo.org/)
is the European network for the Virtual Observatory (VO). It started
in 2001, and currently is composed by 5 national partners (AstroGrid,
GAVO, OV-France, SVO, Vobs.it) and one supranational organisation
(ESA).

The Euro-VO has been funded by the European Commission with several
joint projects. Currently, ESCAPE (The European Science Cluster of
Astronomy and Particle Physics ESFRI Research Infrastructures;
https://projectescape.eu) is the project that is funding the Euro-VO
since 2019. This project has received funds from the European Union's
Horizon 2020 research and innovation programme under the Grant
Agreement no 824064.

ESCAPE is a large project focus on bringing together three scientific
continuities which are actually connected: astronomy, astroparticles,
and particle physics. ESCAPE trays to identify synergies in the domain
of open data management and data systems. On this sense, the Euro-VO
is supporting the implementation of VO standards into the European
Open Science Cloud.

\section{The VO schools}

One important aspect of the Euro-VO network is the support on the
utilization of VO tools and services by the European Astronomical
community. One way to do this is by mean of VO schools, that Euro-VO
has been offering for many years. Since 2009 we have had 10 VO schools
at European level. That is a school each one or two years. In
addition, the Spanish Virtual Observatory group regularly organizes VO
schools for the Spanish community.

These schools are mainly focus on early-career astronomers (PhD
students and early post-doc), although any member of the Astronomical
community can assist. One important aspect of the schools is the
individual attention. Thus, the numbers of students is limited to 30-50
per edition, depending on the venue, and the number of tutors is among
10 and 12. So typically, each tutor attends 3 o 4 students.

In the past, the VO schools were in person. But due to the pandemic
situation, in February 2021 we afforded the challenge of moving it to
an on-line format. The selected platforms were Zoom, for guiding the
live-on sessions, and Slack, for holding an off-line forum for
questions and discussions. The number of participants was similar to
the previous editions. The effect of moving to virtual school was
positive after all. In fact, we noticed more interaction between
students and tutors than in the previous in person schools. The school
was very positive evaluated by the students.

This gratifying experience has encouraged us to plan in February 2022
the next ESCAPE VO school with an hybrid format, allowing the
participation both in person and on-line, if the pandemic situation
permits.

\articlefiguretwo{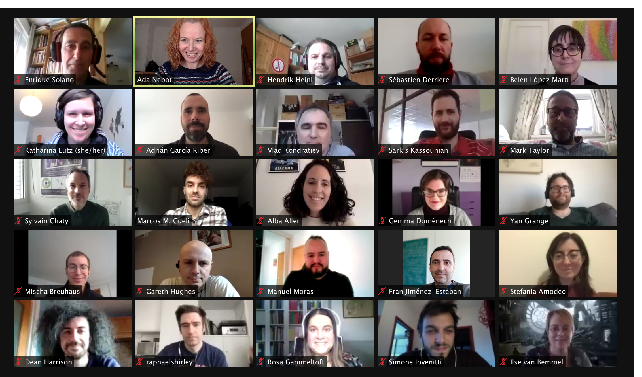}{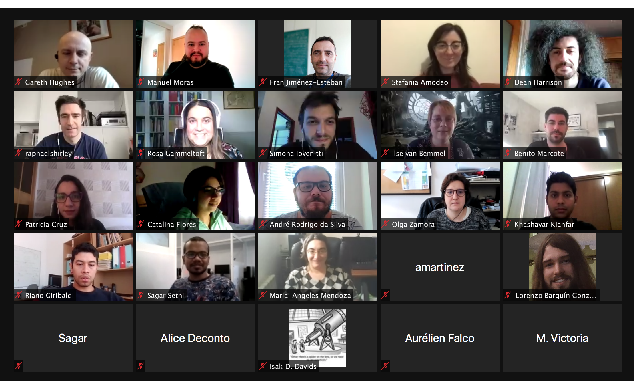}{ex_fig2}{Group picture taken during our last on-line Euro-VO school.}

\section{Our approach}

The approach we follow is focused on reaching the two main goals of
the school: i) expose researchers to the currently available VO tools and
services, so they can learn about the functionalities and
possibilities that the VO offers, with the final objective that they
use them for their daily research afterwards; ii) gather their
feedbacks on the VO tools, services, and the school itself.

In order to reach the first goal, the schools have two main
ingredients. The first is the hand-on sessions, which actually occupy
the majority of the time. Typically, five or six of these sessions are
hold during the school. They are based on real scientific cases, and
they are supported by a detailed written guide which were prepared in
advance by the tutors. The sessions are driven by one tutor, while 2
or 3 other tutors resolve any individual issue arising to any
students.

The second important ingredient is the participant use-cases. Before
the schools, participants are encouraged to bring use-cases that may
be of interest for their own research. These cases are tackled during
the school with the direct guidance of the tutors, and using the new
learned techniques.

Finally, to reach the second main goal, retrieve feedback from the
students, we survey them on the VO tools and services, and on the
tutorials that were used during the schools. In general, students give
high rate to the school and to the VO tools and services. In fact,
they normally express their intention of using VO in their future
research, confirming that we successfully reach the first main goal.

\section{Training material}

As has been explained above, we use tutorials for the hand-on
sessions. The tutorials are based on real science cases which are
developed step-by-step following a VO approach. The tutorials are very
detailed, so students can follow them with very few or even no help at
all.

These tutorials are saved in a public repository where any interested
person can access, download them, and trainee in VO by following the
exercises. Some of these repositories are listed below:
\begin{itemize}
\item https://www.euro-vo.org/scientific-tutorials/
\item https://www.asterics2020.eu/tutorials/overview
\item https://svo.cab.inta-csic.es/docs/index.php?pagename=Meetings
\item http://vo-for-education.oats.inaf.it/eng\_download.html
\end{itemize}

{\bf Acknowledgements.} We acknowledge the support of ESCAPE (European
Science Cluster of Astronomy and Particle Physics ESFRI Research
Infrastructures) funded by the EU Horizon 2020 research and innovation
program under the Grant Agreement n.824064.



\end{document}